\documentclass[sn-nature]{sn-jnl}

\usepackage{graphicx}%
\usepackage{multirow}%
\usepackage{amsmath,amssymb,amsfonts}%
\usepackage{amsthm}%
\usepackage{mathrsfs}%
\usepackage[title]{appendix}%
\usepackage{xcolor}%
\usepackage{textcomp}%
\usepackage{manyfoot}%
\usepackage{booktabs}%
\usepackage{algorithm}%
\usepackage{algorithmicx}%
\usepackage{algpseudocode}%
\usepackage{listings}%
\makeatletter
\newcommand*{\rom}[1]{\expandafter\@slowromancap\romannumeral #1@}
\makeatother

\theoremstyle{thmstyleone}%
\theoremstyle{thmstyletwo}%

\theoremstyle{thmstylethree}%

\begin{document}

\title[Article Title]{Avoiding lateral mode leakage in thin film lithium niobate waveguides for the generation of spectrally pure photons at telecom wavelengths}

\author{\fnm{Muskan} \sur{Arora}}\email{idz218420@iitd.ac.in}

\author{\fnm{Pranav} \sur{Chokkara}}\email{idz218513@iitd.ac.in}

\author*{\fnm{Jasleen} \sur{Lugani}}\email{jasleen@sense.iitd.ac.in}

\affil{\orgdiv{Centre of Sensors, Instrumentation and Cyber Physical System Engineering (SeNSE)}, \orgname{Indian Institute of Technology Delhi}, \orgaddress{\street{Hauz Khas}, \city{New Delhi}, \postcode{110016}, \state{State}, \country{India}}}

\abstract{Photonic integrated optical components, notably straight waveguides, serve as pivotal elements for on-chip generation and manipulation of quantum states of light. 
In this work, we focus on optimizing waveguides based on lithium niobate on insulator (LNOI) to generate photon pairs at telecom wavelength using spontaneous parametric down-conversion (SPDC). Specifically, we 
investigate lateral leakage for all possible SPDC processes involving type 0, type I and type II phase matching conditions in an X-cut lithium niobate waveguide and provide a recipe to avoid leakage loss for the interacting photons. 
 Furthermore, focusing on type II phase matching, we engineer the waveguide in the single mode regime such that it also satisfies group index matching for generating spectrally pure single photons with high purity (99.33\%). We also address fabrication imperfections of the optimized design and found that the spectral purity of the generated photons is robust to fabrication errors. This work serves as a tutorial for the appropriate selection of morphological parameters to obtain lossless, single mode LNOI waveguides for building linear optical circuits and photon pair generation at telecom wavelengths using desired phase-matching conditions.}




\maketitle

\section{Introduction}
Quantum states of light such as single photons serve as the cornerstone for various quantum optical tasks that involve interference between multiple, indistinguishable and spectrally pure photons \citep{kok2007linear,spring2013boson,couteau2023applications}. 
One of the most popular methods to generate such quantum states is through spontaneous parametric down-conversion (SPDC) process using second order non-linear optical medium, which could be either in bulk crystals or integrated waveguides configuration. In this regard, thin film lithium niobate on insulator (LNOI) \citep{saravi2021lithium,vazimali2022applications} has emerged as a strong integrated photonic platform which leverages on the versatile material properties of lithium niobate (LN) and nanoscale waveguide fabrication and is being widely explored for quantum optical tasks \citep{boes2018status,krasnokutska2018ultra}. 

In the last few years, LNOI has gathered a lot of attention for the generation of quantum light sources \citep{ebers2022flexible,duan2020generation}. On-chip generation of broadband, entangled photon pairs has been reported with high brightness \citep{javid2021ultrabroadband,zhao2020high}. Generation of spectrally uncorrelated photon pairs in fully etched silica clad and partially etched MgO doped LNOI straight waveguides has been demonstrated recently \citep{kumar2022group,xin2022spectrally}. 
In \citep{briggs2021simultaneous}, simultaneous type I and type II phase matched SPDC processes is realized in integrated lithium niobate waveguide by using higher order modes. Apart from these, fundamental optical components such as polarization beam splitters (PBS) with high extinction ratios\citep{gong2017optimal,wu2022lithium}, high-performance tunable narrowband grating filters \citep{prencipe2021tunable} and low-loss  multimode interference (MMI) couplers \citep{li2023low} have been realized on LNOI. 
Most of the above-mentioned schemes are based on partially etched rib waveguide geometry in LNOI and involve multiple higher order modes which in certain circumstances can lead to cross talk and mode interference, impacting the device efficiency.  
However, in the context of linear optical computing, single-mode waveguides are required for building fundamental optical components like beam splitters and Mach-Zehnder interferometers etc.

Also, similar to silicon-on-insulator \citep{kakihara2009generalized,nguyen2019lateral,webster2007width}, rib waveguides in LNOI also suffer from lateral mode leakage which is a prominent source of loss in such waveguides, especially when shallow etched \citep{yu2022study,kang2023shallow}.
Lateral leakage occurs when the guided mode of the waveguide gets phase-matched with the orthogonally polarized slab mode and leaks into the slab structure. This happens when the effective index of the waveguide mode (TM/TE) becomes less than the mode index of the slab mode of the opposite polarization (TE/TM) \citep{boes2021integrated}. 
This leakage is inherent in shallow etched rib waveguides and the only way to avoid them is to choose appropriate waveguide parameters. \\ Lateral leakage in LNOI waveguides has been reported earlier \citep{yu2022study, saitoh2011design,kang2023shallow} and investigated for its effects on non-linear optical processes such as second harmonic generation \citep{boes2021efficient} and for designing high extinction filters \citep{boes2021integrated}.
However, a detailed analysis on the morphology of LNOI waveguides to avoid lateral leakage loss and producing photon pairs with the desired properties using type 0, type I and type II SPDC processes is not available.
 In this work, we address this issue and perform a comprehensive analysis for the lateral mode leakage in LNOI waveguides and provide a recipe to choose waveguide dimensions appropriately to avoid leakage losses for all the interacting wavelengths in type 0, type I, and type II SPDC processes. During optimization, we also ensure the single-mode operation for the downconverted photons at telecom wavelengths. We then focus specifically on type II phase matching condition and engineer the waveguide such that it gives rise to spectrally pure photons with high purity (99.33\%) at telecom wavelength. We also check for fabrication tolerances of the optimized design and find that the proposed waveguide geometry is robust towards fabrication errors in terms of the spectral purity of the generated photons.

\section{Theory of SPDC process for generating spectrally pure photons}

 We start by briefly discussing the basic theory of SPDC process. The two-photon output state resulting from an SPDC process can be written as \citep{jin2013widely,lugani2020spectrally}  

\begin{equation}
    |\psi(\omega_s, \omega_i)\rangle = \iint f(\omega_s , \omega_i) a^{\dag}_{s} a^{\dag}_{i} |0_s\rangle |0_i\rangle d\omega_s d\omega_i
\end{equation}
Here, $f(\omega_s , \omega_i)$ is joint spectral amplitude (JSA)  which is equal to the product of pump envelope function (PEF), $\alpha(\omega_s + \omega_i)$ and phase matching function (PMF), $\phi(\omega_s, \omega_i)$ :
 \begin{equation}
    f(\omega_s , \omega_i) = \alpha(\omega_s + \omega_i) * \phi(\omega_s, \omega_i) 
    \label{}
\end{equation} 
where 
 \begin{equation}
    \alpha(\omega_s + \omega_i) = exp{\left[-\frac{(\omega_s + \omega_i - \omega_p)^2}{\sigma_p^2}\right]}
    \label{Eq. (2)}
\end{equation}  
  and                            
 \begin{equation}
    \phi(\omega_s, \omega_i) = sinc\left[\Delta k(\omega_s, \omega_i)\frac{L}{2}\right] exp{\left[i\Delta k(\omega_s, \omega_i)\frac{L}{2}\right]}
\end{equation}  \textit{$\omega_{p(s,i)}$}  is the angular frequency of pump (signal, idler) photon, \textit{$\sigma_p$} is the pump bandwidth, \textit{L} is the length of the waveguide and \textit{$\Delta k$} is the phase mismatch given by:
\begin{equation}
 \Delta k = k_p - k_s - k_i - \frac{2\pi}{\Lambda}
\end{equation}
  \textit{$k_{p(s,i)}$} are the wavevectors of pump (signal, idler) photons and \textit{$\Lambda$} is the poling period required to satisfy the phase matching condition. \\

The shape of the JSA $f(\omega_s , \omega_i)$ gives an important information about the spectral correlations between the downconverted photons \citep{zielnicki2018joint}. The slope of PEF is always negative due to energy conservation (Eq.\ref{Eq. (2)}), whereas the slope of PMF can be controlled through the dispersion engineering of the non-linear medium and can thus be used to tailor spectral correlations between signal and idler photons. 
The slope of PMF can be obtained by considering Taylor series expansion \citep{jin2013widely,mosley2007generation} of \textit{$\Delta k$}, which can further be expressed in the terms of group indices of the pump ($n_{g}^{p}$), signal ($n_{g}^{s}$) and idler photons ($n_{g}^{i}$) as \citep{kumar2022group,xin2022spectrally} :

\begin{equation}
    tan(\theta) = -\frac{n_{g}^{p} - n_{g}^{s}}{n_{g}^{p} - n_{g}^{i}}
\end{equation}
where $\theta$ is the angle made by PMF with the horizontal (signal frequency) axis. \\

 In order to create spectrally uncorrelated photons, $\theta$ should satisfy $0^{\circ} \leq \theta \leq 90^{\circ}$, according to which the group index ($n_{g}^{p}$) of the pump photon should either lie between the group index of signal ($n_{g}^{s}$) and of idler ($n_{g}^{i}$) photon or should be equal to one of them i.e,
\begin{equation}
 n_{g}^{s} \leq n_{g}^{p} \leq n_{g}^{i}  \text{ or } n_{g}^{i} \leq n_{g}^{p} \leq n_{g}^{s} 
 \label{Eq. (7)}
\end{equation}
This, in principle, is possible by tailoring the dispersion profile of the waveguide by engineering its geometry. The resulting two photon state will be spectrally uncorrelated by a proper choice of pump bandwidth. 
The JSA then becomes factorizable for signal and idler frequencies and can be written as  
\begin{equation}
      |\psi(\omega_s, \omega_i)\rangle= |\xi_i(\omega_s) \rangle |\zeta_i(\omega_i) \rangle.
      \end{equation}
Spectral purity of the output photons can be quantified by performing singular value decomposition (SVD) of the obtained JSA \citep{mosley2007generation}. 

\section{Waveguide optimization}
 \begin{figure}[b]
\begin{center}
        {\includegraphics[width=4.5cm]{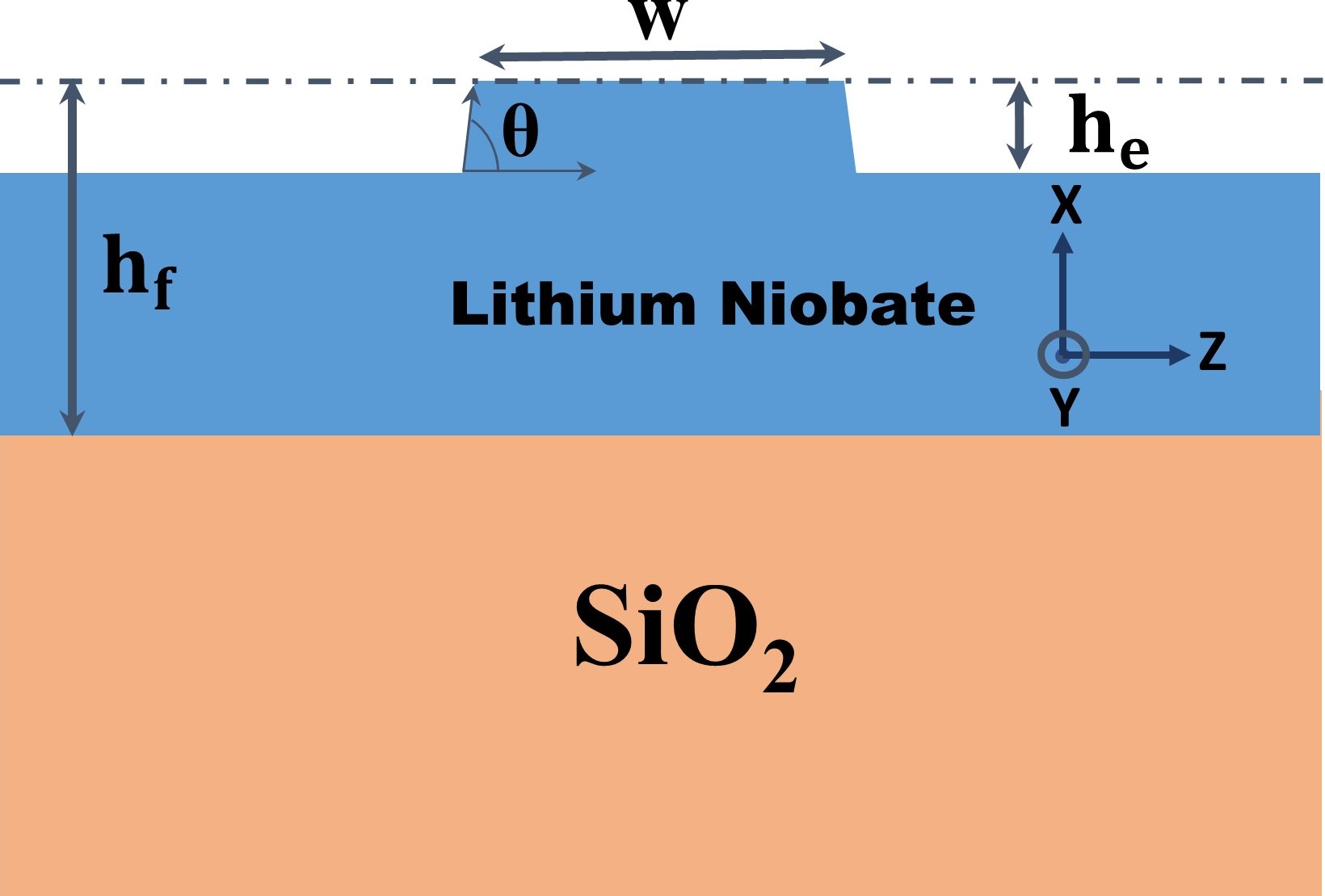} }
        \caption{Schematic of the cross-section of LNOI waveguide with top width (\textit{w}), etching depth (\textit{${h_e}$}), thickness (\textit{${h_f}$}) and sidewall angle (\textit{$\theta$}).}
        \label{FIG. 1}
         \end{center}
        \end{figure}

\begin{figure*}
\begin{center}
        {\includegraphics[width=14 cm]{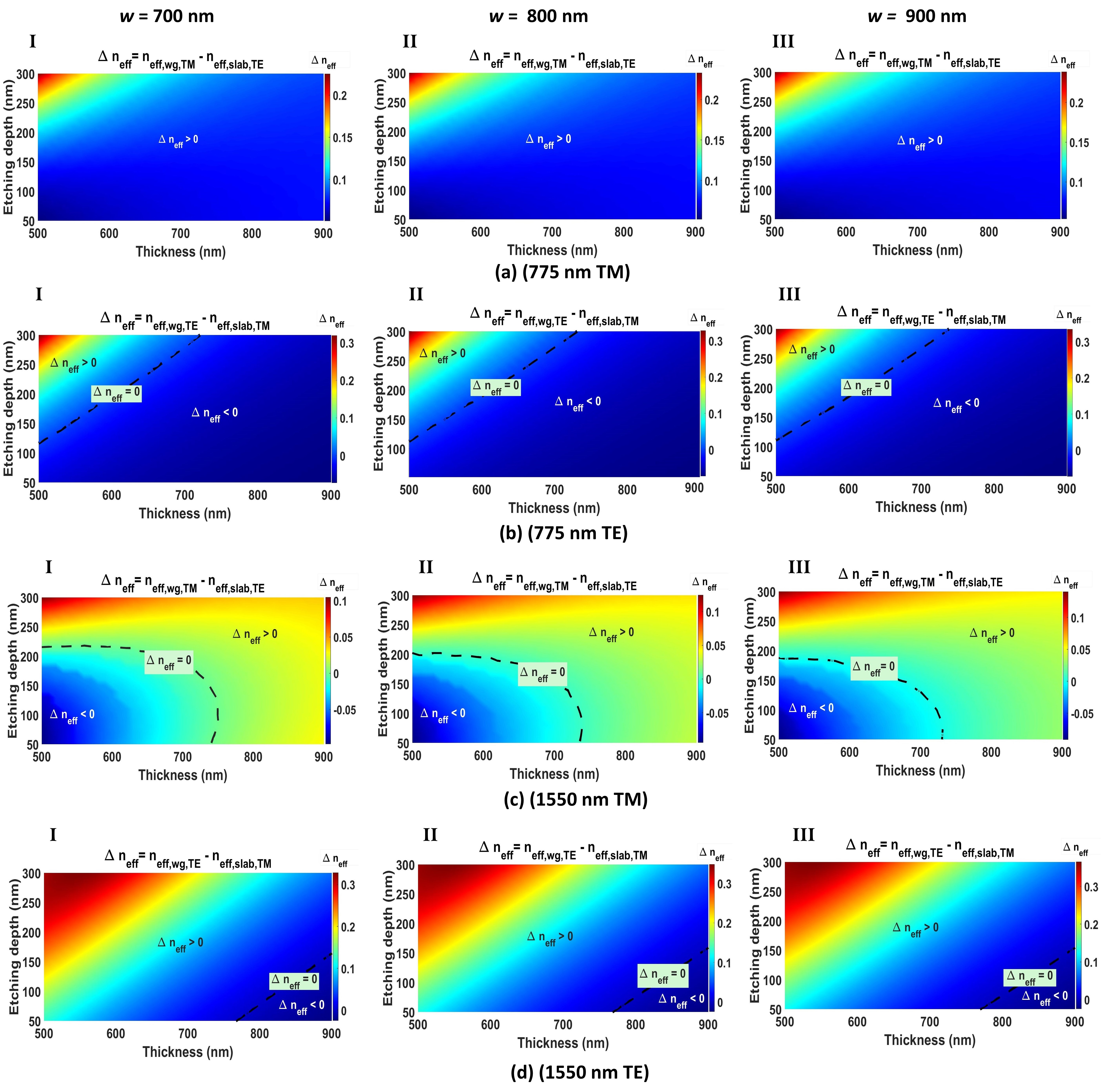} }%
       \caption{Variation of $\Delta n_{eff}$ with etching depth (\textit{${h_e}$}) and thickness (\textit{${h_f}$}) for modes at (\rom{1}) 775 nm TM, (\rom{2}) 775 nm TE,  (\rom{3}) 1550 nm TM and  (\rom{4}) 1550 nm TE for three top widths( \textit{w}) (a) 700 nm, (b) 800 nm and (c) 900 nm.}%
         \label{FIG. 2}
         \end{center}
     \end{figure*}

We optimize an X-cut LNOI waveguide whose typical cross-section with top width (\textit{w}), etching depth (\textit{${h_e}$}), total thickness  (\textit{${h_f}$}) and sidewall angle (\textit{$\theta$}) is shown in \hyperref[FIG. 1]{Fig.1}. 
The waveguide optimization process is centered around three goals: (a) to investigate lateral leakage for all possible, i.e., type 0,  type I and type II phase matched SPDC processes for photon pair generation at telecom wavelengths, (b) to ensure single mode operation for the downconverted photons, and (c) to find a single-mode waveguide geometry that is devoid of leaky modes and generates spectrally pure photons at telecom wavelengths. \\
\subsection{Optimization of waveguide for photon pair generation using type 0, type I and type II SPDC processes }
For understanding lateral leakage as a  function of the waveguide geometry, 
we consider all types of phase matching that is possible in an X-cut LN waveguide which includes (a) type 0 phase matching in which pump, signal and idler all have the
same (TE) polarization, (b) type I phase matching in which signal and idler, both have
same polarization (TM) which is opposite to that of the pump (TE), and (c) type II phase matching
in which the signal and idler have orthogonal polarizations (TM and TE). These are described below with the corresponding non-linear coefficients:\\

(a) \textbf{Type 0}:   \ \  Pump (775 nm, TE polarized) $\rightarrow$  Signal (1550 nm, TE polarized) + Idler (1550 nm, TE polarized),  \  \    through $d_{33}$ coefficient  

(b) \textbf{Type I}:  \ \   Pump (775 nm, TE polarized) $\rightarrow$ Signal (1550 nm, TM polarized) + Idler (1550 nm, TM polarized),  \  \   through $d_{31}$ coefficient 

(c) \textbf{Type II}: \ \ Pump (775 nm, TM polarized) $\rightarrow$  Signal (1550 nm, TM polarized) +  Idler (1550 nm, TE polarized),   \  \      through $d_{31}$ coefficient  \\

In order to investigate lateral leakage, we compute a parameter called \textit{$\Delta n_{eff}$} which is the difference between the effective indices of the waveguide mode and orthogonally polarized slab mode as a function of different waveguide parameters (\textit{w},\textit{${h_e}$},\textit{${h_f}$}). As described above, waveguide modes leak into the slab structure if the effective index of the waveguide mode (TM/TE) becomes less than the mode index of the slab mode of the opposite polarization (TE/TM) \citep{boes2021integrated}. \hyperref[FIG. 2]{Fig.{2}} shows the variation of  $\Delta n_{eff}$ values as a function of different etching depths (\textit{${h_e}$}= 50 nm to 300 nm) and LN thicknesses (\textit{${h_f}$}= 500 nm to 900 nm) for pump (775 nm), signal (1550 nm) and idler (1550 nm) photons, for different top widths (\textit{${w}$})  of the waveguide ranging from 700 nm to 900 nm. The side wall angle is kept at a constant value of 80$^{\circ}$. These simulations are performed using the finite difference eigenmode (FDE) solver in Ansys Lumerical software. 

We first start by discussing the case of type II phase matching, which involves modes at 775 nm TM (pump), 1550 nm TM (signal) and 1550 nm TE (idler).
It can be seen from \hyperref[FIG. 2]{(Fig. {2}a(\rom{1}-\rom{3}))} that the pump mode exhibits positive values of $\Delta n_{eff}$ for all the considered waveguide parameters, indicating a very low probability of lateral mode leakage for a wide range of waveguide dimensions for pump photon.
However,  $\Delta n_{eff}$ varies from negative to positive values for signal and idler modes for different waveguide geometries \hyperref[FIG. 2]{(Fig.{2})}. For the regions, where $\Delta n_{eff}$ $\leq$ 0, there is high probability of mode leakage and thus high leakage losses \citep{kang2023shallow,boes2021integrated}. For signal modes  \hyperref[FIG. 2]{(Fig. {2}c(\rom{1}-\rom{3}))}, we discuss the case of width 700 nm \hyperref[FIG. 2]{(Fig. {2}c(\rom{1}))} and width 800 nm \hyperref[FIG. 2]{(Fig. {2}c(\rom{2}))} .
We observe that for all the thicknesses which are less than 740 nm, mode leakage can occur when etching depth is close to or less than 220 nm as the value of $\Delta n_{eff}$ is negative for this region. However, for thicknesses greater than 740 nm, the probability of leakage loss is very less, for the entire range of etching depths considered as $\Delta n_{eff}$ $>$ 0 in this region. The crossover from negative to positive values of $\Delta n_{eff}$ is indicated by the (black) dashed lines.
A similar variation can be seen in case of width 900 nm \hyperref[FIG. 2]{(Fig. {2}c(\rom{3}))}. 

For idler modes \hyperref[FIG. 2]{(Fig. {2}d(\rom{1}-\rom{3}))}, in case of width 700 nm \hyperref[FIG. 2]{(Fig. {2}d(\rom{1}))}, for LN thicknesses greater than 766 nm, there is a high probability of leakage if etching depths are less than 166 nm. However, for etching depths greater than 166 nm, probability of leakage is low for a wide range of thicknesses (500 nm- 900 nm). 
Similar variation is observed in case of other widths \hyperref[FIG. 2]{(Fig. {2}d(\rom{2}))} and \hyperref[FIG. 2]{(Fig. {2}d(\rom{3}))}.
 It must be noted that in order to eliminate lateral mode leakage for the considered SPDC processes, geometries for which  $\Delta n_{eff}$ $\leq$ 0 for any of the interacting wavelengths should be avoided.\\ 

We now consider  the case of type I and type 0 phase matching which involves pump at 775 nm in TE polarization.  
Unlike for TM polarized pump discussed above, we find that in case of TE polarized pump \hyperref[FIG. 2]{(Fig. {2}b(\rom{1}-\rom{3}))} , $\Delta n_{eff}$ $\le$ 0 for certain  geometries in considered range of waveguide dimensions. We discuss the case of width 800 nm  \hyperref[FIG. 2]{(Fig. {2}b(\rom{2}))}. 
It can be seen that for LN thicknesses  lying between 500 nm and 730 nm, the probability of leakage is very high if etching depth is less than 120 nm (as $\Delta n_{eff}$ $\le$ 0). Also, as the thickness increases from 500 nm to 730 nm, the etching depth required to avoid also leakage loss also increases.
 For thickness 500 nm, the value of etching depth should be atleast 120 nm while for thicknesses 600 nm and 700 nm, the etching should be atleast 200 nm and 280 nm respectively. 
 However, for the thicknesses greater than 730 nm, the etching depth should be greater than 300 nm. 
 Similar analysis can be done in the case of other top widths 700 nm \hyperref[FIG. 2]{(Fig. {2}b(\rom{1}))} and 900 nm  \hyperref[FIG. 2]{(Fig. {2}b(\rom{3}))}. \\

  Using the above analysis, we can choose the waveguide dimensions appropriately such that the waveguide remains leakage free for all the interacting photons at pump, signal and idler wavelengths for a given phase matched SPDC process. For instance, for type 0 and type I phase matching  which involves pump at 775 nm TE polarized, we could select a waveguide of thickness 600 nm, top width 800 nm, etching depth 300 nm and side wall angle 80$^{\circ}$, which is non-leaky for all the wavelengths and is also single mode for the downconverted photons (at 1550 nm). Since the waveguide is largely leakage free for TM polarized pump, we could choose lower etching depths for type II SPDC process, for which the waveguide is single mode for all the interacting wavelengths. The geometry is then further engineered and fine tuned to get the desired output state as described below for the case of spectrally pure photons.

\section{Generation of spectrally uncorrelated photons at telecom wavelength using type II SPDC process}
\begin{figure}
\begin{center}
        {\includegraphics[width=8cm]{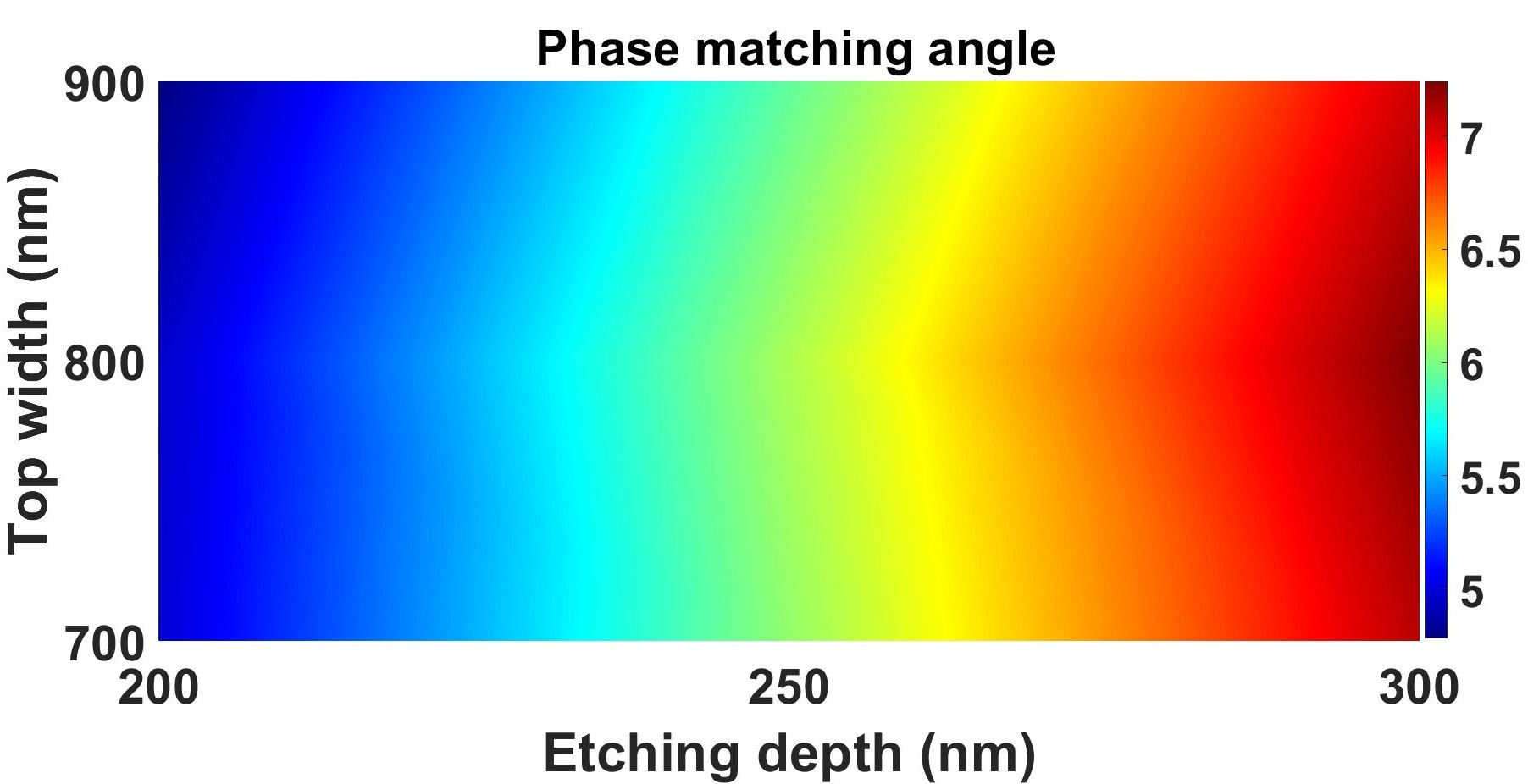} }
        \caption{Variation of phase matching angle with etching depth (\textit{${h_e}$}) and top width (\textit{w}) of the waveguide.}
        \label{FIG. 3}
         \end{center}
        \end{figure}

The goal of this paper is also to generate spectrally uncorrelated photons at telecom wavelengths.
For this, we specifically focus on type II SPDC process which can lead to spectrally pure photons at 1550 nm. As described in Section 2, for obtaining spectrally uncorrelated photon pairs in an SPDC process,  the group indices of the interacting photons should satisfy the relation given in (Eq.\ref{Eq. (7)}), and phase matching angle ($\theta$) should lie between $0^{\circ}$  and $90^{\circ}$.
To explore the group index matching condition as a function of waveguide parameters, we scan the variation of phase matching angle \hyperref[FIG. 3]{(Fig. {3})} as a function of etching depth and top width of the waveguide at the constant thickness (\textit{${h_f}$})  and sidewall angle (\textit{$\theta$}) of  800 nm and 80 $^{\circ}$ respectively. The above  parameters (\textit{${h_f}$}) and (\textit{$\theta$}) are chosen based on the analysis of \hyperref[FIG. 2]{Fig.{2}} to avoid lateral leakage loss.
It can be seen from \hyperref[FIG. 3]{(Fig. {3})} that the minimum values of phase matching angle (less than 5 $^{\circ}$) can be achieved for the etching depth of 200 nm for all the considered widths. \\

\begin{figure}[h]
        \centering
       {\includegraphics[width=10cm]{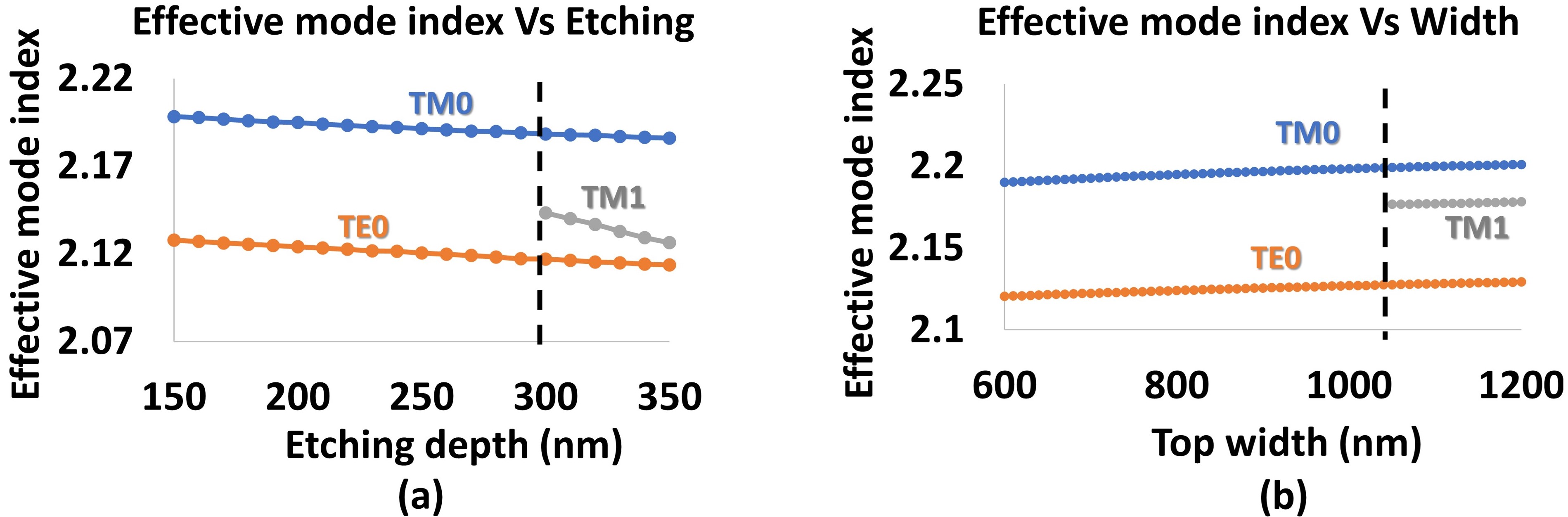} }%
        \caption{Variation of effective mode index of pump mode (775 nm) with (a) etching depth ({${h_e}$}) and (b) top width (w) of the waveguide .}%
        \label{FIG. 4}%
        \end{figure}

 In order to find the dimensions of the waveguide which lies in a single mode regime for type II SPDC process, we investigate the mode structure of the waveguide for the geometries by computing the effective mode indices ($n_{eff}$) of supported modes at 775 nm. \hyperref[FIG. 4]{(Fig.{4}a)} shows the variation of $n_{eff}$ of the modes as a function of etching depth (\textit{${h_e}$}) varying from 150 nm to 350 nm (at \textit{${h_f}$}= 800 nm, \textit{${w}$}= 800 nm). Note that if a waveguide is single mode at a lower wavelength for a specific configuration, it has to be single mode at a higher wavelength as well for the same configuration. As can be seen \hyperref[FIG. 4]{(Fig.{4}a)}, the waveguide is single mode for a maximum etching depth of 300 nm and thereafter it starts supporting higher order modes (TM1). Likewise, \hyperref[FIG. 4]{(Fig.{4}b)} shows the variation of $n_{eff}$ as a function of top width (\textit{${w}$}) ranging from 600 nm to 1200 nm (at \textit{${h_f}$}= 800 nm, \textit{${h_e}$}= 200 nm). It can be seen that the waveguide remains single-moded for a top width of 1050 nm, after that, it starts to show higher-order modes (TM1).
 \\       
All the simulation results of \hyperref[FIG. 2]{(Fig.{2})} , \hyperref[FIG. 3]{(Fig.{3})} and \hyperref[FIG. 4]{(Fig.{4})} collectively provide a recipe for choosing a  waveguide geometry for type II SPDC process which is devoid of lateral mode leakage, single mode for all the interacting wavelengths and also satisfies the group index matching condition, required for producing spectrally uncorrelated photon pairs. \\

\begin{figure*}
        \begin{center}
        \includegraphics[width = 13 cm]{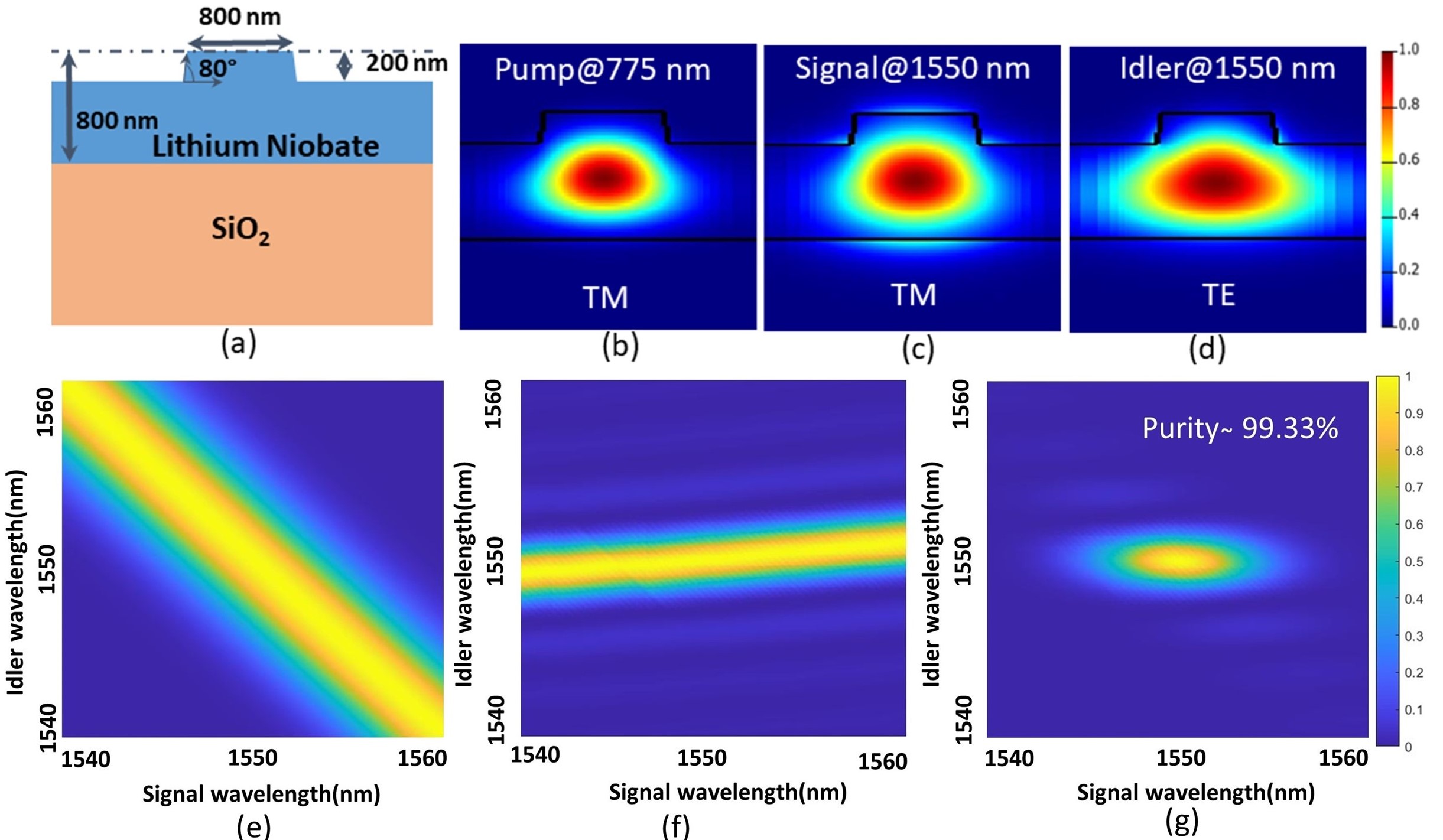} 
        \caption{(a) Schematic of cross-section of LNOI waveguide; Electric filed intensity profiles of (b) pump, (c) signal and (d) idler ; (e) Pump envelope function (PEF); (f) Phase matching function (PMF); (g) Joint spectral intensity (JSI) function}
        \label{Fig. 5}
         \end{center}
        \end{figure*}  
         \begin{figure*}[b]
        \begin{center}
        \includegraphics[width = 6 cm]{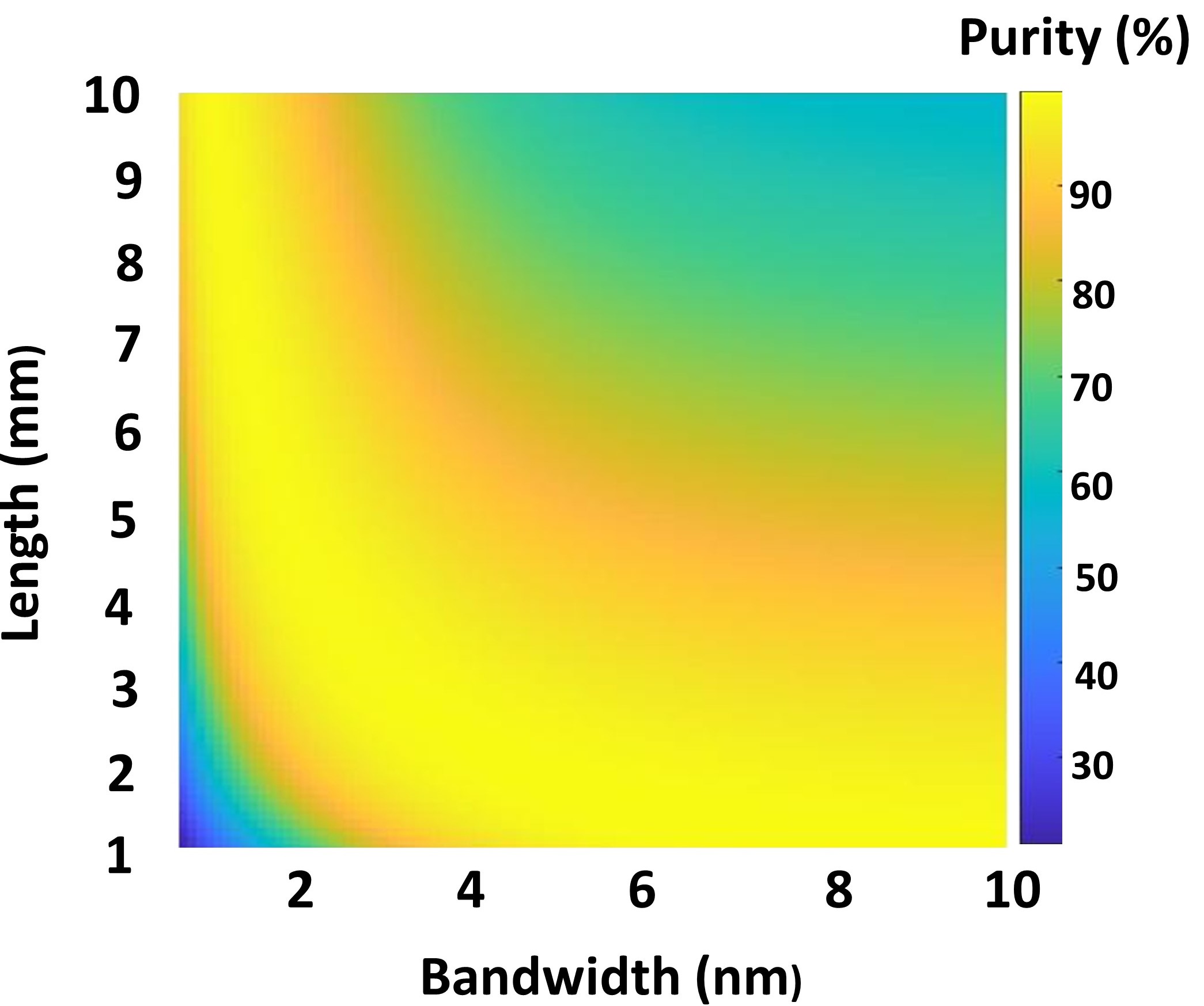} 
        \caption{Variation of spectral purity of the generated photons with pump bandwidth and length of the waveguide }
        \label{Fig. 6}
         \end{center}
        \end{figure*}  

\subsection{Joint Spectral Intensity (JSI) of the output state}
Based on our analysis of waveguide optimization, we consider a geometry that is free of any leakage loss and is also single mode at all the interaction wavelengths ( 775 nm and 1550 nm) and  is shown in \hyperref[Fig. 5]{Fig. {5}(a)} . It has LN thickness of 800 nm, top width of 800 nm, etching depth 200 nm, and sidewall angle 80$^{\circ}$. The electric field intensity profiles for pump, idler and signal photons are shown in \hyperref[Fig. 5]{Fig. {5}(b-d)}. The required poling period is 3.54 microns and the length of the waveguide is chosen to be 5 mm. For this geometry, we obtain the values of group indices of the pump ($n_{g}^{p}$), idler ($n_{g}^{i}$), and signal ($n_{g}^{s}$) to be 2.4276, 2.2567, 2.4425 respectively which satisfies the group index matching condition ($n_{g}^{p}$ $\approx$ $n_{g}^{s}$).       
The PEF and PMF are plotted in \hyperref[Fig. 5]{(Fig.{5}(e,f))}. The JSI ($|f(\omega_s , \omega_i)|^2$) which is a measurable quantity in experiments is plotted in \hyperref[Fig. 5]{(Fig.{5}(g))}. The spectral purity of the biphoton state is computed by performing the SVD of JSI and is found to be 99.33$\%$ for a pump bandwidth of 1.5 nm.
 
Since the purity of the generated photon pairs is limited by the pump bandwidth and length of the waveguide \textit{L}, we also compute and plot spectral purity as a function of these parameters \hyperref[Fig. 6]{(Fig. 6)}, confirming that the generated photons remain spectrally uncorrelated for a wide range of waveguide lengths and pump bandwidths. \\

\section{Fabrication Tolerances}
\begin{figure*}
        \begin{center}
        \includegraphics[width=13.5cm]{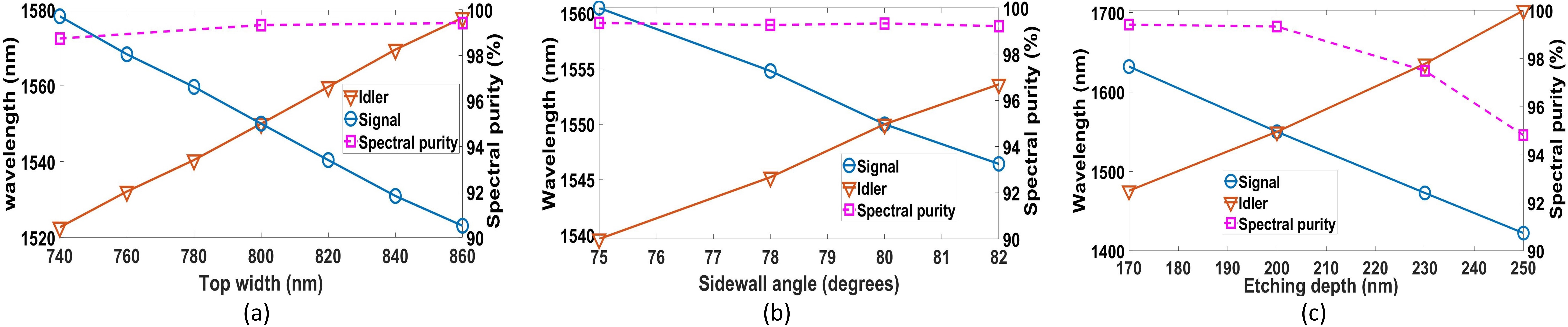} 
        \caption{Variation of phase matched wavelengths and spectral purities as a function of (a) Top width (\textit{w}) (b) Sidewall angle (\textit{$\theta$}) and (c) Etching depth (\textit{${h_e}$}) }
        \label{Fig. 7}
         \end{center}
        \end{figure*}

We further check the fabrication tolerances of our waveguide design by examining the variation of phase-matched wavelengths and corresponding spectral purities of the generated two-photon state. We consider the deviation from the ideal design parameters with respect to top width, sidewall angle and etching depth \hyperref[Fig. 7]{(Fig.{7})} of the waveguide.

We observe that with a deviation of approximately 10 nm in the top width \hyperref[Fig. 7]{(Fig.{7}a)} of the waveguide, the signal and idler show a maximum shift of 5 nm on each side of the desired phase-matched wavelength (1550 nm). However, spectral purities of the generated photons are maintained at about 99$\%$ \hyperref[Fig. 7]{(Fig.{7}a)} even with a deviation of $(\pm 60 nm)$ in the designed top width. Further, the deviation of the side wall \hyperref[Fig. 7]{(Fig.{7}b)} angle by 5$^{\circ}$ results in a maximum shift of approximately 10 nm on each side (signal and idler) from 1550 nm. Again, the purity \hyperref[Fig. 7]{(Fig.{7}b)}  at 75$^{\circ}$ is 99.35$\%$, while at 82 $^{\circ}$, it is 99.21$\%$. However, a deviation of 10 nm in the etching depth \hyperref[Fig. 7]{(Fig.{7}c)} causes a significant shift of approximately 25 nm in each side for signal and idler wavelengths. A change of etching depth from 200 nm to 230 nm drops the spectral purity \hyperref[Fig. 7]{(Fig.{7}c)} from 99.33$\%$ to 97.49$\%$, while for etching depth of 250 nm, purity is around 94.81$\%$ for the shifted phase matched wavelengths.
This indicates that the waveguide is robust to the fabrication deviations from the desired parameters for waveguide top width and sidewall angle but a better control of etching depth is required for maintaining the desired characteristics. \\


\section{Conclusion}

In conclusion, we present a thorough analysis for the optimization of LNOI waveguide geometry for the generation of photon pairs at telecom wavelengths. We provide a recipe for choosing waveguide parameters appropriately to avoid lateral mode leakage for all possible phase-matching conditions (type 0 , type I and type II) in an X cut lithium niobate waveguide. In addition, we also explore the parameter regime in which the waveguide is single mode for the downcoverted photons at 1550 nm for all the processes. Our work serves as a tutorial for choosing the waveguide parameters appropriately for the photon pair generation at telecom wavelengths using desired SPDC phase matched process. Furthermore, based on our detalied anaylsis, specifically focusing on type II SPDC process, we find a  loss-free single-mode waveguide geometry that satisfies group index matching condition for generating cross-polarized spectrally pure photons with high spectral purity (99.33$\%$). Our chosen geometry is also robust to fabrication imperfections in terms of spectral purity of the generated photons. We believe that the reported work on optimization of LNOI waveguides with respect to lateral leakage and mode structure will serve as a  useful guide to develop low loss optical components on this platform for both classical and quantum optical tasks including photon pair generation with desired properties. 

\backmatter

\bmhead{Acknowledgments}
This work is supported by a startup research grant by SERB, India (SRG/2022/001759) and PhD scholarships by Ministry of Human Resource Development (MHRD).





\bibliography{sn-article}

\end{document}